\documentstyle[epsf]{article}
\newcommand{\gsim}{\mbox{\raisebox{-.3em}{$\stackrel{>}{\sim}$}}}
\newcommand{\lsim}{\mbox{\raisebox{-.3em}{$\stackrel{<}{\sim}$}}}

\renewcommand{\cite}[1]{\ref{#1}}

\newcommand{\beq}{\begin{equation}}
\newcommand{\eeq}{\end{equation}}
\newcommand{\beqa}{\begin{eqnarray}}
\newcommand{\eeqa}{\end{eqnarray}}
\newcommand{\bpr}{\begin{problem}}
\newcommand{\epr}{\end{problem}}
\newcommand{\bcent}{\begin{center}}
\newcommand{\ecent}{\end{center}}
\newcommand{\bfig}{\begin{figure}}
\newcommand{\efig}{\end{figure}}
\newcommand{\bpc}{\begin{picture}}
\newcommand{\epc}{\end{picture}}
\newcommand{\barr}{\begin{array}}
\newcommand{\earr}{\end{array}}
\newcommand{\bitm}{\begin{itemize}}
\newcommand{\eitm}{\end{itemize}}
\newcommand{\bright}{\begin{flushright}}
\newcommand{\eright}{\end{flushright}}
\newcommand{\bminip}{\begin{minipage}}
\newcommand{\eminip}{\end{minipage}}

\newcommand{\lmd}{\lambda}

\newcommand{\reflef}{(\ref}
\newcommand{\MP}{M_{\rm P}}

\begin{document}
\baselineskip=0.6cm

\bcent
{\Large\bf Time-variability of the fine-structure constant expected from the Oklo constraint and the QSO absorption lines}\\[.6em]
Yasunori Fujii\\[.5em]
Advanced Research Institute for Science and Engineering, \\
Waseda University, Shinjuku, Tokyo 169-8555, Japan\\[.5em]
{\bf Abstract}\\[.5em]
\bminip{13cm}
\baselineskip=.4cm
The data from the QSO absorption lines indicating a nonzero time-variability of the fine-structure constant has been re-analyzed on the basis of a ``damped-oscillator" fit, as motivated by the same type of behavior of a scalar field, dilaton,  which mimics a cosmological constant to understand the accelerating universe.  We find nearly as good fit to the latest data as the simple weighted mean.  In this way, we offer a way to fit the more stringent result from the Oklo phenomenon, as well.
\eminip

\ecent
\baselineskip=.6cm

\section{Introduction}

In this article we suggest a possible way to understand the nonzero time-variability of the fine-structure constant $\alpha$ reported recently from the observed QSO absorption lines [\cite{ww1},\cite{ww2}], in conformity with the more stringent constraint from the Oklo phenomenon [\cite{shly}-\cite{yfetal}].  Interesting enough this consistency is derived from the suspected behavior of the scalar field required to fit the accelerating universe [\cite{randp}], another striking finding in today's cosmology.  The rest of this Section will be devoted to a brief account on the background as well as the summary of the result.

Attempts to probe $\Delta\alpha$ for the difference from today's value of the fine-structure constant $\alpha$ based on the calculated relativistic corrections of many transitions is now summarized [\cite{ww2}] to give $\Delta\alpha /\alpha =(-0.54\pm 0.12)\times 10^{-5}$ for the redshift range $0.2-3.7$, or the fractional look-back time $0.2-0.9$.  The estimate according to the weighted mean of the 128 data points, which we accept in the present analysis, may be referred to as the 1-parameter (horizontal straight-line) fit,  for the convenience of comparison with our own fit in terms of a ``damped-oscillator" model with three parameters, as will be explained shortly.

The 1-parameter fit in this sense appears to be confronted by the constraint smaller by 2 or 3 orders of magnitude as was established by the measured isotopic abundance of $^{149}{\rm Sm}$ in the remnant of Oklo natural reactors.   This result, which, despite apparent complications of nuclear physics, is relatively free from uncertainties thanks to an exceptionally low-lying resonance, is confirmed by careful re-analyses of Ref. [\cite{shly}], yielding either of the upper bounds $|\Delta\alpha /\alpha | \lsim 10^{-7} $ [\cite{dd}] or $\lsim 10^{-8}$  [\cite{yfetal}], or the nonzero value $\sim 10^{-7}$ [\cite{yfetal}] still not completely ruled out.  The small ratio of the resonance energy 97.3 meV to $\sim$MeV that characterizes typical energy scale of nuclear phenomena makes it also plausible to derive the result on $\Delta\alpha/\alpha$ without entering into details of the strong-interaction coupling constant [\cite{yfetal}].

No immediate conflict is encountered, however,  because the fractional look-back time around 0.14 for the Oklo phenomenon is outside the range for the QSO observation.  The small time-separation between them still seems to indicate a rather sharp time-dependence of $\alpha$, obviously requiring a theoretical interpretation.  Note also that it imposes a much more stringent constraint than the results from such low-$z$ phenomena, like H{\footnotesize I} 21cm and molecular rotational transitions [\cite{D98}].

It was argued that the issue can be related to the accelerating universe which is accounted for by a cosmological constant with $\Omega_\Lambda \sim 0.7$ [\cite{randp}] by allowing the time-dependent speed of light [\cite{BM1}].  Also shown was the potential importance of spatial inhomogeneity of time variation of $\alpha$, concluding that the phenomena taking place on the Earth's surface might be affected by the local matter density more than those in the galactic clouds [\cite{BT}].  We pursue here instead a more conventional approach by making use of somewhat nontrivial time-dependence suggested by the behavior of the scalar field supposed to result in an effective cosmological constant.  Possible sharp change of the scalar field has also been discussed in connection either with the running nature of its coupling constant [\cite{wett}], or with a classical field, like ours but motivated with different potentials [\cite{anchor},\cite{gardner}].

Our detailed analysis of the accelerating universe in terms of the scalar-tensor theory [\cite{yf5},\cite{cup}] indicates a particular type of behavior of the scalar field, sometimes called a dilaton $\sigma$, which stays nearly constant that mimics a cosmological constant.  A closer look shows further that this ``plateau" is superimposed with a small vibration, reminiscent of a damped oscillation as a function of the cosmological time, as will be explained briefly in Section 2.  We combined this with a relation $\Delta\alpha/\alpha \propto \Delta\sigma$, which we derive based on a specific model, but might be viewed as an extension of the previous results [\cite{bek},\cite{fon}].  We try to see if this oscillation pattern can be fitted to the observed $\Delta\alpha/\alpha$ with a zero of the oscillation identified with the small value constrained by the Oklo phenomenon [\cite{bjnporto}].

In view of still unavoidable uncertainties in selecting a unique solution of the cosmological equations, we attempt a phenomenological approach by assuming the damped oscillation behavior of $\Delta\alpha/\alpha$ with three arbitrary coefficients which are determined to fit the QSO observation.  We show that this 3-parameter fit to the QSO result can be nearly as good as the 1-parameter fit in [\cite{ww1},\cite{ww2}], fitting the Oklo data as well.  Notice that the 1-parameter fit, if extended to the Oklo time, would yield an unacceptably large value of $\chi^2$.

One might still wonder why we are content with  $\chi_{\rm red}^2 = 1.09$ for the former with more adjustable parameters is not significantly smaller than $\chi_{\rm red}^2 = 1.06$ for the latter.  In reply, we are content, at this moment, with the result that our fit motivated by a theoretical prejudice survives a realistic test.  This ``success" will be followed by  future attempts to scrutinize the solutions for the accelerating universe.  Admitting that the result is far from final, we still believe it to serve as a useful first step toward better understanding the issue in a wider perspective.

In Section 2, we begin with describing the current theoretical model following [\cite{yf5},\cite{cup}].  Section 3 offers our 3-parameter fit and presenting the result, to be compared with the starting cosmological solution.   The final Section 4 is devoted to concluding remarks, including a discussion on the constraint proposed recently on the Re-Os decay [\cite{olive}].

\section{Theory of the cosmological constant and the time-dependent fine-structure constant}

Today's cosmological constant problem has two faces [\cite{cup}]: Why is $\Lambda_{\rm obs}$ so small compared with the theoretical expectation by 120 orders of magnitude?  Why is it still nonzero?

The scalar-tensor theory provides an acceptable answer to the first question.  The theory seems to deserve two renewed interests; possible origin of the scalar field in the dilaton of string theory, and a way to implement the scenario of a decaying cosmological constant. One of the crucial theoretical ingredients is in choosing a correct conformal frame (CF).  String theory is formulated naturally in the J(ordan) CF, featuring a ``nonminimal coupling" of the dilaton field (though in higher dimensions).  We also accept that, almost in any version of unification models, a cosmological ``constant" $\Lambda$ is present in the J frame, namely in the ``string CF."

A ``physical" CF is to be chosen, on the other hand, according to the two requirements; i) keeping the solution of the cosmological equations with $\Lambda$ included from deviating too much off the standard cosmology, ii) resulting in the masses of matter fields time-independent.  The second condition is imposed because our clocks and meter sticks are made basically of the particle masses.  A solution to this non-trivial problem is offered by assuming a ``scale-invariance" in 4 dimensions except for the $\Lambda$ term, finding the E(instein) CF with the Einstein-Hilbert term as physical.  Particle masses emerge nonzero as the invariance is broken spontaneously. The $\Lambda$ term in the J frame is now converted to an exponential potential of the scalar field $\sigma$. Unlike in the more phenomenological approach with the quintessence models, no inverse-power potential is expected.  The energy density of $\sigma$, dark energy, falls off like $t^{-2}$, thus implementing the scenario of a decaying (effective) cosmological constant [\cite{declmd}].

The scalar field $\sigma$ remains decoupled from matter in the classical limit, but is brought to matter-coupling due to the interactions among matter fields, {\em \`{a} la} quantum anomaly.  According to the simplified QCD calculation, the coupling constant to the quark is obtained; $g_{\sigma {\rm q}}= \zeta (5\alpha_s/\pi)\approx 0.3 \zeta$, where $\zeta$ is a constant of the order one, $\alpha_s \approx 0.2$ being the QCD analog of the fine-structure constant, while $\MP^{-1}= 8\pi G$ has been suppressed.  The fact that the matter coupling depends on the strong-interaction constant signals composition-dependence of the force mediated by $\sigma$, leading to WEP violation.

The coupling constant to nucleons is also computed; $g_{\sigma {\rm N}}\sim 0.019 \zeta$.  The relatively small value is a consequence of the smallness of the quark-mass component in the nucleon mass.  The $\sigma$ eventually acquires a nonzero mass, as a pseudo Nambu-Goldstone boson, resulting in a medium-ranged force.  An exact estimate of the force-range $\lmd$ is not easy, but is certainly shorter than distances that characterize solar-system experiments, which therefore no longer constrain the parameters of the $\sigma$ force.

The parameter $\alpha_5$ in the non-Newtonian potential between two nucleons $-(Gm_{\rm N}^2/r)(1+\alpha_5 e^{-r/\lmd})$ [\cite{yfnn}] is given by $\alpha_5 = 2g^2_{\sigma {\rm N}}\approx 0.72\times 10^{-3}\zeta^2$, being consistent with the composition-independent experiments roughly for $\lmd \lsim 1{\rm m}$ [\cite{fsb}] including the recent submillimeter experiment [\cite{eotwash}].  Results of composition-dependent experiments may be analyzed by replacing $\alpha_5 \rightarrow \alpha_5 q_i q_j$ with the ``charge" $q_i$ for the $i$-th substance. For a free-fall experiment on the Earth, we find the relative acceleration difference $(\Delta g)_{ij,{\rm E}}/g \sim \alpha_5 (\Delta q)_{ij}\lmd \times 4\times 10^{-8}$, where $\lmd$  is measured in units of meters.  In view of still present uncertainties in $(\Delta q)_{ij}$, we borrow tentatively from the vector fifth-force model, $|(\Delta q)_{ij}| \sim 2\times 10^{-3}$ [\cite{fsb}].  The available observational upper bounds $|(\Delta g)_{ij,{\rm E}}/g|\sim 10^{-12}$ is respected if $\lmd \lsim 10{\rm m}$, thus giving an idea how large the expected WEP violation can be.

The second face, or question, requires some deviation from the smooth behavior of the energy density $\rho_\sigma$. As one of the simplest approaches along this direction, we introduced another scalar field, $\chi$, assumed to have a specific interaction with $\sigma$.  A typical cosmological solution for $k=0$ is shown in Fig. 1 around the present epoch (taken from Fig. 5.8 of [\cite{cup}]).  The energy density $\rho_s$ of the dark energy $\sigma, \chi$ behaves like a plateau, acting as a cosmological constant.  As a consequence of the dynamics of the scalar fields, the plateau comes to a crossing with the ordinary matter density.  Nearly in coincidence with this, the scale factor exhibits a ``mini-inflation," that fits the observed ``acceleration," also with concomitant sudden changes of $\sigma$ and $\chi$.

\noindent
\bminip{7.4cm}
\baselineskip = 0.4cm
\epsfxsize=7.2cm
\epsffile{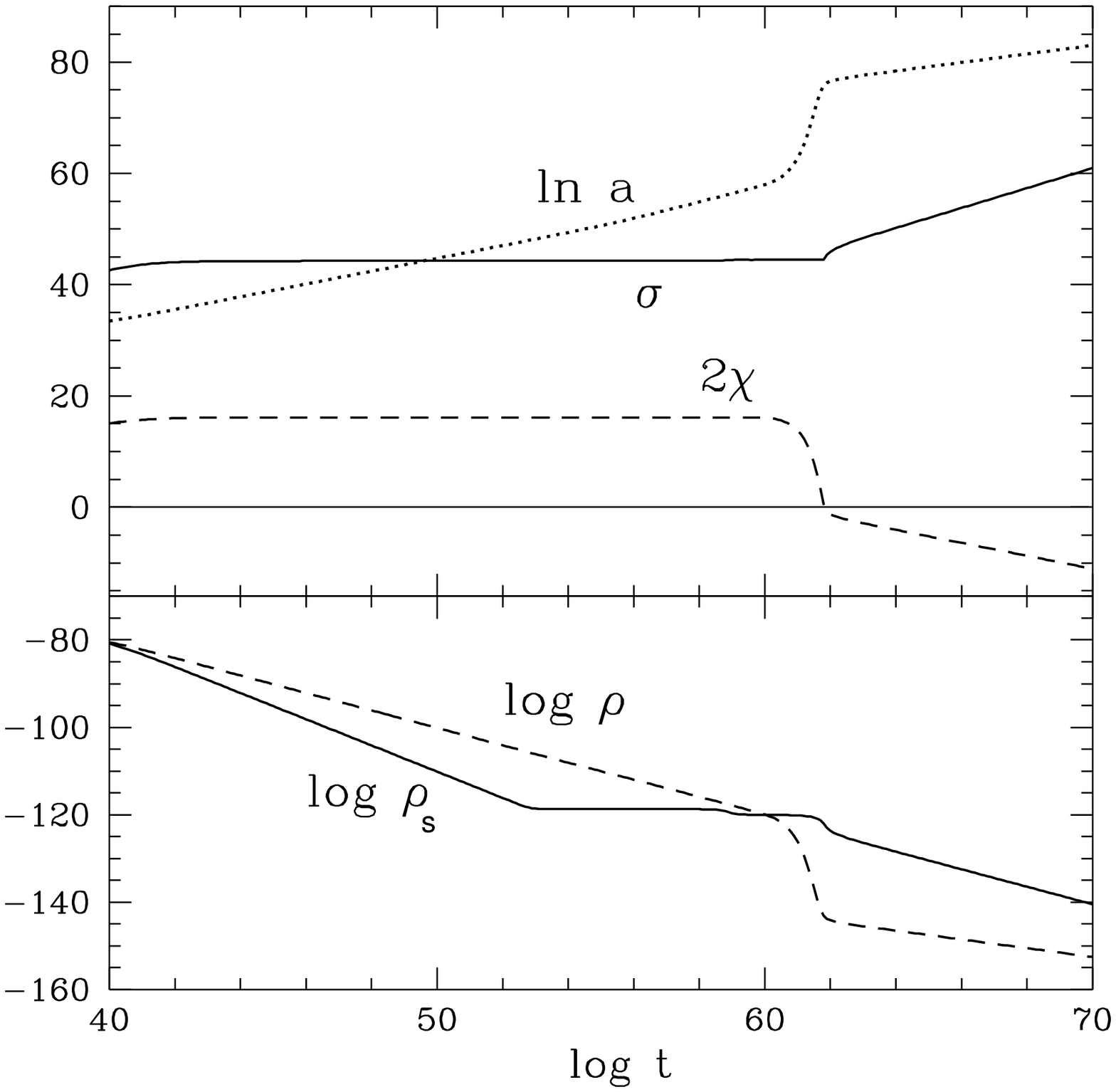}
\noindent
{\small Figure 1: An example of the Friedmann solution with $k=0$ of the cosmological equations of the two-scalar model, taken from Fig. 5.8 of [\cite{cup}].  On the horizontal axis, $\log t\approx 60$ corresponds to the present epoch in the (reduced) Planckian unit system of $c=\hbar = \MP (= (8\pi G)^{-1/2})=1$.  In the lower panel, $\rho$ and $\rho_s$ for the ordinary matter density and the energy density of the dark energy, namely for $\sigma$ and $\chi$, respectively are shown.

}
\mbox{}\\[1.1em]
\eminip
\hspace{8mm}
\bminip{7.4cm}
\baselineskip = 0.4cm
\epsfxsize=7.2cm
\epsffile{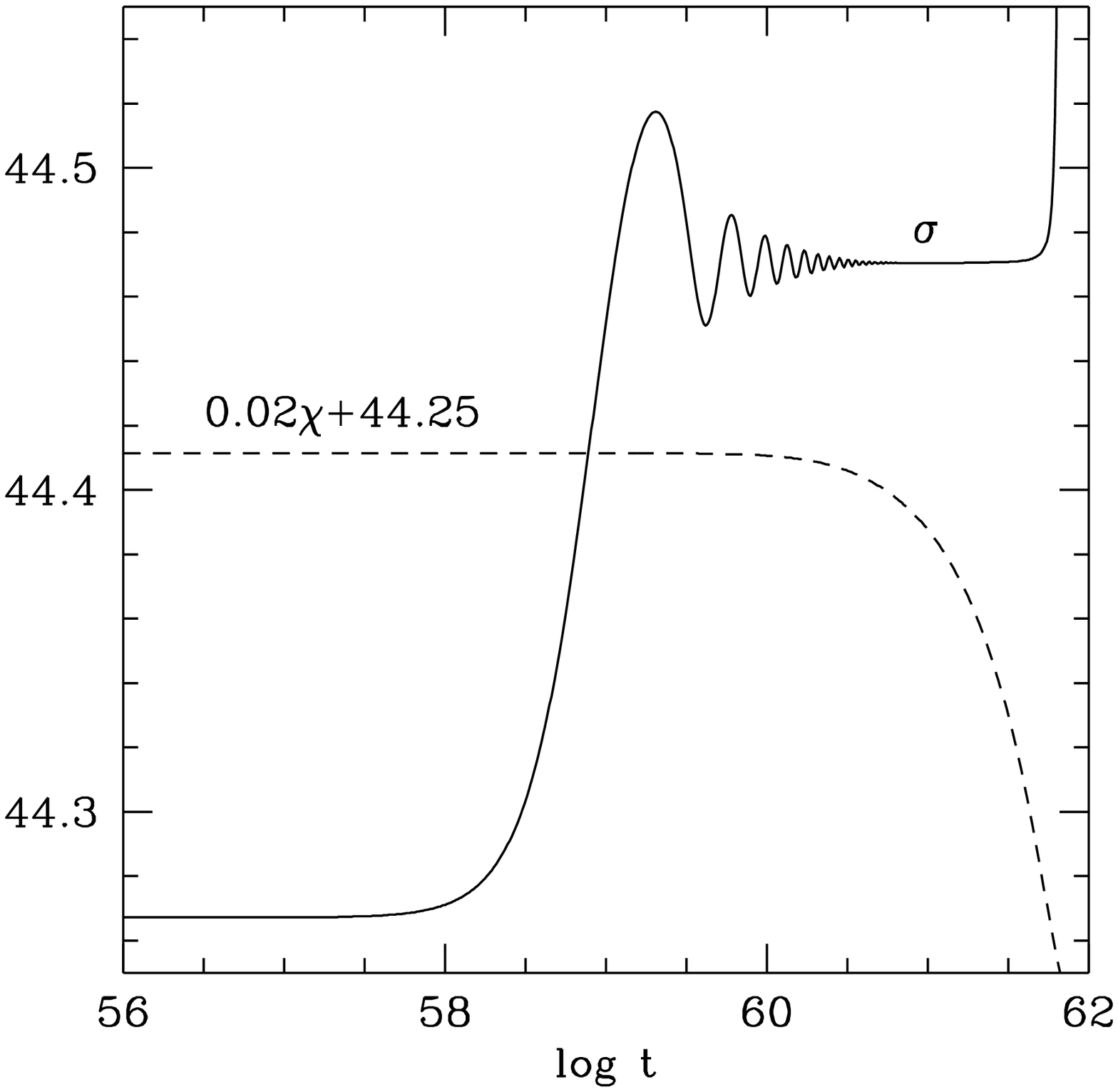}
\noindent
{\small 
Figure 2: Magnified view of $\sigma$ (solid) and $0.02 \chi+44.25$ 
(dashed) {\em vs} $\log t$ near the present epoch.  Note that before magnification, this is part of the plot (previous figure) in which $\sigma$, after the period of plateau behavior resulting in a ``mini-inflation" of the scale factor, shoots up ``suddenly."  }
\mbox{}\\[1.2em]
\eminip

Particularly interesting is found if the apparently simple jump of $\sigma$, as shown in Fig. 1, is vertically magnified greatly, by 330 times, for example, as shown in Fig. 2.  This unexpected damped-oscillator-like behavior plays a key role in bringing about an accelerating universe.  The oscillation is ``small," however.  We recognize it only after the plot is magnified considerably than what we need to analyze how the cosmological acceleration takes place.  Different cosmological solutions leading to nearly identical results  may be differentiated only by looking at other phenomena, including $\Delta\alpha/\alpha$, among others.

We derived the relation (6.194) of [\cite{cup}] through the QED calculation;
\beq
\frac{\Delta\alpha}{\alpha}= {\cal Z}\frac{\alpha}{2\pi}\zeta\Delta\sigma,
\label{thalp-1}
\eeq
where ${\cal Z}$ counts the effective number of charged fields that contribute to the loop.   We consider this result only as an example of the expected relation on the idea that a possible variation of the coupling constant is due to the changing scalar  field, as we may obtain directly from string theory, though uncertainties seem unavoidable for deriving results applicable directly to 4 dimensions.  In the following we take a more phenomenological attitude that the pattern of $\sigma (t)$ is described by a simplified damped oscillation with the parameters determined to fit the observation of $\Delta\alpha/\alpha$.

\section{Analysis of the data}

We begin with Fig. 2 showing $\sigma$ and $\chi$  as functions of $\log t$, which is a natural choice of the time variable in solving the cosmological equation.  The function $\sigma$ is readily converted to that of the fractional look-back time $u= (t_0 -t)/t_0$, with $t_0$ the present age of the universe [\cite{ww1},\cite{ww2}], essentially in the backward time direction.  From a phenomenological view suggested above, we assume the purely damped oscillation behavior for $y=(\Delta\alpha/\alpha) \times 10^5$;
\beq
y = ae^{bu} \sin \left( 2\pi \frac{u-u_{\rm oklo}}{T} \right),
\label{thalp-2}
\eeq
where $u_{\rm oklo}$ is chosen to be 0.142 corresponding to the mean $(1.95 \pm 0.05) {\rm Gy}$ of the time of the Oklo phenomenon.   The natural reactors are believed to have lasted somewhat shorter than $10^6$ years, corresponding to a width in the $u$-direction as small as $\lsim 10^{-4}$.  According to a positive (negative) $a$, $y(u)$  increases (decreases) at $u=u_{\rm oklo}$.  We should choose $b>0$.

We notice that choosing \reflef{thalp-2}) results in $y(0)=a\sin \left( -2\pi u_{\rm oklo}/T \right)$ which may not vanish in general, in contradiction with the natural condition $y(0)=0$ to be obeyed by definition.  As the simplest remedy we may add an offset constant to the right-hand side of \reflef{thalp-2}), but calling for another parameter.  Improving the analysis in this direction will be made in the future, however, by emphasizing the behavior in the ranges close immediately to the QSO range, for the moment.

The data set to be fitted is taken from the ``fiducial sample" (Fig. 8) of [\cite{ww2}] of 128 points.  A 1-parameter fit results in
\beq
y = -0.54 \pm 0.12,
\label{thalp-3}
\eeq
with $\chi^2_{\rm red} =\chi/127 =1.06$, which would have been 1.24 if we were to try to fit the data by the null result $y=0$.  We emphasize that the error bar at the Oklo point is invisibly small ($\lsim 10^{-2}$) compared with other error bars.  It then follows that the 1-parameter fit if extended to $u_{\rm oklo}$  would have resulted in $\chi^2_{\rm red}$ as large as the order of $\gsim 10^4$.  In order for the result to be sensible, we must bend the straight line \reflef{thalp-3}) to pass the point $u_{\rm oklo}$ on the $u$-axis, but hopefully with a theoretical rationale.

\noindent
\bminip{7.3cm}
\baselineskip=0.4cm
\bcent
\begin{tabular}{|| c | c c c c  ||}\hline\hline
Solution & $b$ & $a$ & $T$  & $\chi^2_{\rm red}$ \\ \hline
${\rm A}_+$ & 2.4 & \hspace{1em}0.151 & 0.714 & 1.09 \\
${\rm B}_+$ & 0.0$^*$ & \hspace{1em}0.350 & 0.275  & 1.22 \\
${\rm C}_+$ & 3.2 & \hspace{1em}0.041 & 0.182 & 1.22 \\
${\rm A}_-$ & 4.0$^*$ & $-0.035$ & 0.451 & 1.20 \\
${\rm B}_-$ & 2.2  & $-0.085$ & 0.218 & 1.21 \\
${\rm C}_-$ & 3.0  & $-0.037$ & 0.158 & 1.23 \\\hline\hline
\end{tabular}
\ecent
{\small Table 1: Six solutions for minimized $\chi^2$, and their parameters.  The asterisks mean the end-point minima instead of the genuine local minima.}
\eminip
\hspace{8mm}
\bminip{7.3cm}
\baselineskip=0.4cm
\mbox{}\\
\bcent
\begin{tabular}{|| c | c c c c ||}\hline\hline
$b$ & $a$ & $T$ & $\chi^2$ & $\chi^2_{\rm red}$ \\ \hline
0.0 & 0.728 & 0.716 & 136.83 & 1.10 \\
2.4 & 0.151 & 0.714 & 136.19 & 1.09 \\
4.0 & 0.051 & 0.710 & 136.46 & 1.09 \\\hline\hline
\end{tabular}
\ecent
{\small Table 2: Parameters for the local minima for the solution ${\rm A}_+$, corresponding to the three curves in Figs. 3 and 4.  In the second line for $b=2.4$, the values are for the  3-dimensional minimum of $\chi^2$, while other two lines for $b=0.0$ and $b= 4.0$ represent only the 2-dimensional minima listed for comparing $\chi^2$ with the true minimum for $b=2.4$.}
\mbox{}\\[1.5em]
\eminip

Using the behavior of the scalar field for the accelerating universe as a guide, we conveniently consider the range $0\leq b \leq 4$, inside of which we search for sets of parameters $a, b, T$ that minimize $\chi^2$.   The Oklo point is not included in $\chi^2$ because choosing $u_{\rm oklo}$ is meant to fit the data already.  We have found six solutions, with the parameters and $\chi^2_{\rm red} = \chi^2 /125$ summarized in Table 1.  Remarkably, only the solution ${\rm A}_+$ offers $\chi^2_{\rm red}=1.09$ which is small enough nearly comparable with the value 1.06 for the 1-parameter fit \reflef{thalp-3}), while others share larger values $\sim 1.2$ close to 1.24 for the null result mentioned above.  There are yet other solutions with even larger $\chi^2$.  In what follows we focus on ${\rm A}_+$.

We consider $\chi^2$ in 3-dimensional space of $a, b, T$.  For each solution there is a 3-dimensional local minimum, around which there occurs a region of confidence level of 68\%, for example.  For ${\rm A}_+$, the true minimum occurs for $b=2.4, a= 0.151, T= 0.714$, as given by the first line of Table 1, which are also reproduced in the second line of Table 2.  The surface which surrounds the confidence volume may be represented by the three cross sections each of which is a contour in the 2-dimensional $a-T$ plane with constant $b$, as shown by  Fig. 3 and Table 2.  The complete 3-dimensional minimum lies on the plane of $b=2.4$, as marked by a cross inside a contour labeled by $2.4$ in Fig. 3, while the 2-dimensional minima are also shown as dots inside  the other two contours for $b=0.0$ and $b= 4.0$.  The corresponding values of $\chi^2$ are given in Table 2.  The largest contour occurs for $b=0.0$, though the corresponding minimum $\chi^2$ is larger than that of $b=2.4$.  In this connection we further point out that $b=0.0$ is likely outside the practical interest, because we expect that the present epoch is well before the time of sudden changes of the scalar fields, as shown in Fig. 2.

Figure 4 shows how the (solid) curve of ${\rm A}_+$ fits the data.  Also shown are the curves of neighboring fits corresponding to $b=0.0$ (dotted) and $b=4.0$ (dashed), which nearly overlap that of $b=2.4$ in the range of the QSO observation.  There is an indication, however, that the larger $b$ makes it the easier to fit the Oklo result.  In fact choosing $b=8.0 (a= 0.00297, T= 0.712)$ gives the curve to be within $\pm 0.001$, hence satisfying the ``stronger" Oklo constraint $|\Delta\alpha/\alpha| \lsim 10^{-8}$ in the range $u= 0.142 \pm 0.02$. This is made, however, with the cost of somewhat larger $\chi^2_{\rm red}= 1.11$, because the QSO data exhibits quite ``flat" distribution, instead of a sharp change implied by a large $b$.  In fact this flatness makes it unlikely that the smaller value at the Oklo point is largely due to the damping effect, $e^{bu}$, even taking aside that a large $b$ is not favored by the cosmological behavior.

\noindent
\bminip{7.4cm}
\baselineskip = 0.4cm
\epsfxsize=7.2cm
\epsffile{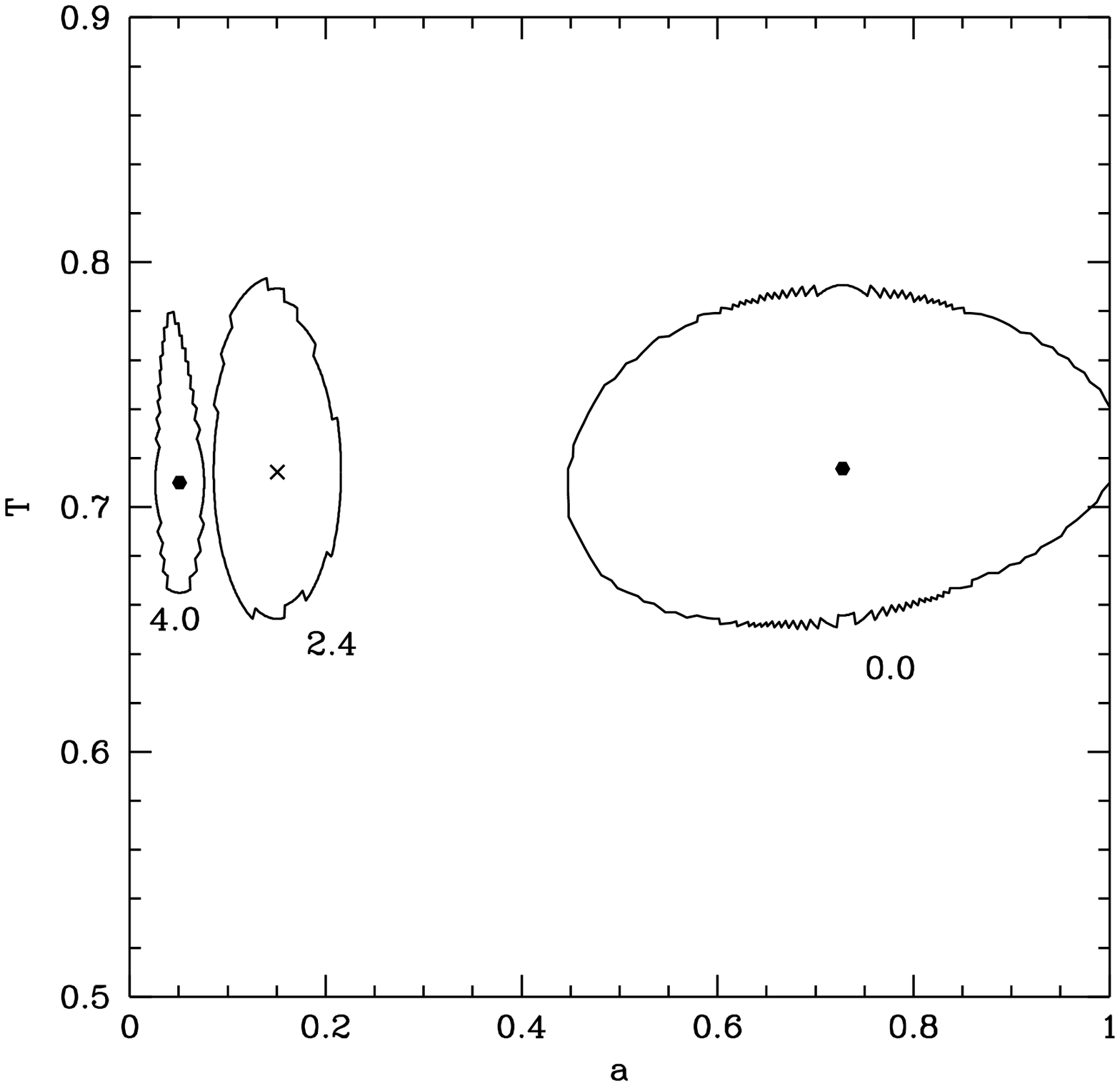}
\noindent
{\small 
Figure 3: Cross sections by $a-T$ planes, labeled by the values of $b$, of the 68\% confidence level of the solution ${\rm A}_+$.  The true 3-dimensional minimum of $\chi^2$ occurs for the point denoted by a cross inside the contour labeled by 2.4, as also given in Table 2.  The points denoted by dots inside other two contours represent only the 2-dimensional minima for each $b$.
}
\mbox{}\\[1.1em]
\eminip
\hspace{8mm}
\bminip{7.4cm}
\baselineskip = 0.4cm
\epsfxsize=7.2cm
\epsffile{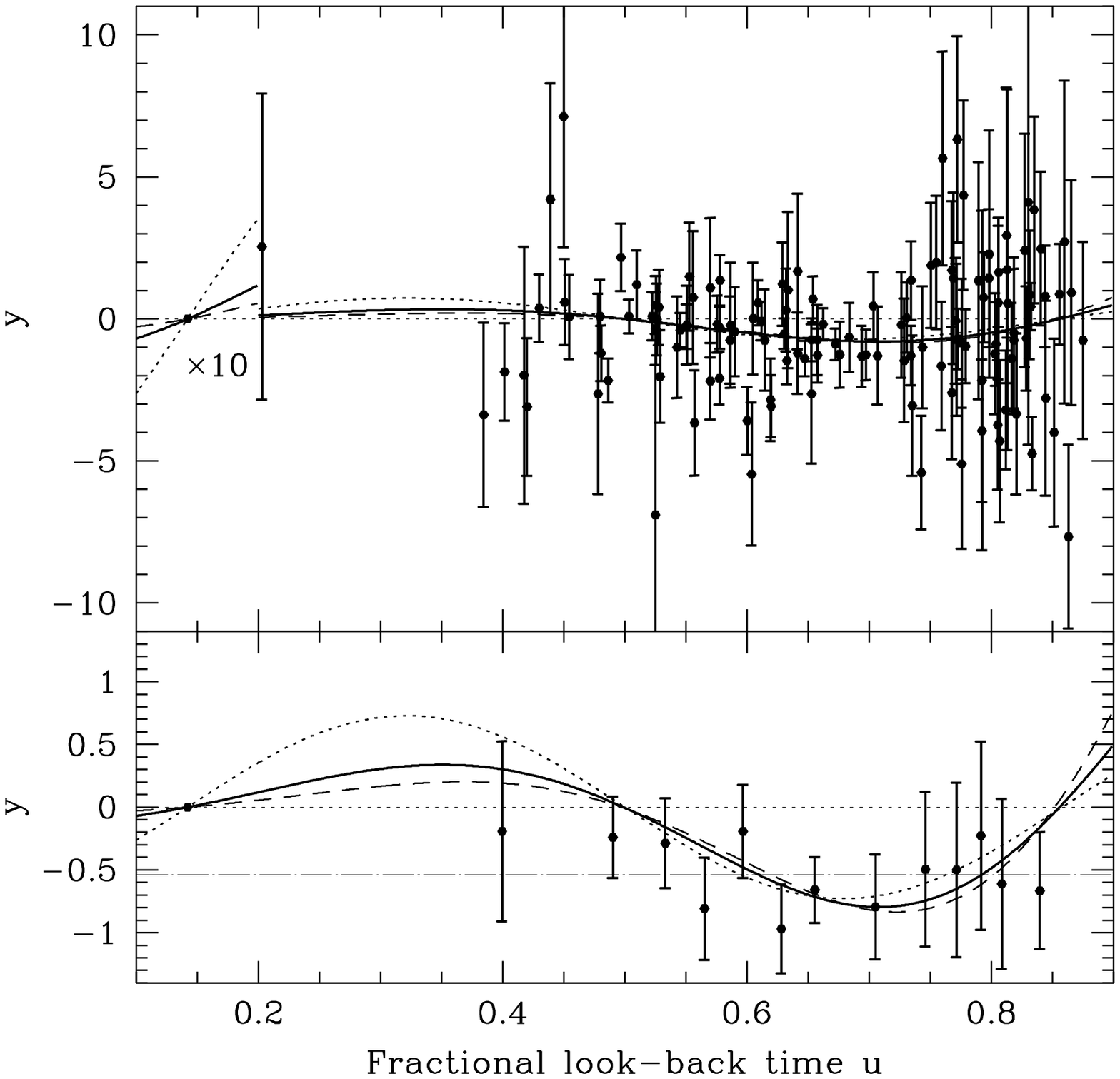}
\noindent
{\small 
Figure 4: Fit of the solution ${\rm A}_+$ (solid).  Similar curves (dotted for $b=0.0$,  dashed for $b=4.0$) are also shown corresponding to Table 2. In the upper panel, the curves for $u<0.2 $ have been 10 times magnified.   In the lower panel (binned data), the (dot-dashed) horizontal straight line is added corresponding to the 1-parameter fit.
}
\mbox{}\\[1.2em]
\eminip

\section{Concluding remarks}

We have shown that the QSO data can be fitted by a damped-oscillator fit nearly equally well as the 1-parameter fit.  Also we can do this by fitting the Oklo result as well.  We have, however, some unsolved problems.

The behavior as shown in Fig. 1 indicates $b\approx 2.5$ and $T\approx 0.22$, compared with $b=2.4$ and $T=0.714$ for the fit ${\rm A}_+$; a near agreement for $b$ but the latter $T$ seems somewhat larger than the former.  Smaller $T$ tends to give larger $\chi^2_{\rm red}$ as we infer from Table 1.  Also the maximum of $|\Delta\sigma|$ around $\log t \approx 59.6$ for $u\approx 0.7$ is $\sim 0.1$, which should be compared with $|y|_{\rm max}\sim 1$ indicated in Fig. 4.  In the cosmological solution in [\cite{cup}] we used the value $\zeta = 1.58$.  Combining these with ${\cal Z} = 5$ from the standard gauge theory, we find that $\Delta\alpha/\alpha$ in ${\rm A}_+$ is smaller than the right-hand side of \reflef{thalp-1}) by  about 2 orders of magnitude.

We add a comment on the $\beta$-decay of $^{187}{\rm Re}$, which is recently claimed to require $|\Delta\alpha/\alpha| \lsim 3\times 10^{-7}$ for the last 4.5 Gys, or $u\leq u_{\rm Re}\approx 0.33$ [\cite{olive}].  The solution ${\rm A}_+$ and its neighbors shown in Table 2 give the values that fail to comply this constraint.  Even the choice of $b=8.0$ as before gives $4.3\times 10^{-7}$, which comes only close to the required bound.  We may reverse the analysis by first identifying the rhenium constraint with a zero of $y$, and finding $(\Delta\alpha/\alpha)_{\rm Oklo}\approx 1.0\times 10^{-7}$.  This might be accepted as the largest possible deviation of the Oklo constraint, with $\chi^2_{\rm red}= 1.09$, but the required $b=8.0$ (also with $a=-0.00366, T= 1.052$) is again quite outside the range allowed from cosmology.

It appears that the rhenium constraint can be made only barely consistent with the flat distribution of the QSO result, at the best.  The same can be a problem for other non-oscillatory behaviors as well.  The function used in [\cite{wett}], for example, features a sharp decrease of $|\Delta\alpha/\alpha|$ from $z=2$ to $z= 0.45$ for the rhenium time to account for the small value.  This entails, in turn, a sharp increase toward the high-$z$ end of the QSO range, agreeing only marginally with the data including not only the average value but also more details, which can be exploited by a $\chi^2$ analysis, for example.  A conclusion of the similar nature will follow also from Fig. 3 of [\cite{anchor}] and Fig. 13 of [\cite{gardner}], for which the estimated $\chi^2_{\rm red}$ appear to remain within an acceptable range, but $|\Delta\alpha/\alpha|$ calculated at the rhenium time turn out somewhat larger than the target value.

In this connection we point out that estimating the parameters of Re-Os isochron accurately is primarily for obtaining the time-averaged decay-rate [\cite{isochron}] eventually to determine the age of meteorites.  Compared with the Oklo phenomenon and the QSO absorption lines, the possible time-variation of the decay-rate, namely of $\alpha$, at certain time has been inferred only by the argument which is less direct and less compelling, as will be elaborated separately [\cite{yfai}].

As a unique feature of the current model, we point out that the behavior of $\sigma$ toward the present epoch depends sensitively on the status at much earlier times.  Figure 5.14 of [\cite{cup}] provides examples to show how small and smooth changes of the initial value may affect whether an oscillation occurs or not, and if it does, how much the amplitude and the frequency vary.  The example shown in Fig. 2 turns out to have a relatively large amplitude.  A much smaller amplitude can be chosen without noticeably affecting the way the universe is accelerated.  Also the interaction between $\sigma$ and $\chi$ was assumed to be the simplest ((5.58) of [\cite{cup}]) to implement the mechanism for the acceleration.  One might even modify this interaction in such a way that fitting the Oklo data as well as meeting the condition $y(0)=0$ can be made more natural.  These details will be the subject of future studies.   We re-emphasize that measuring $\Delta\alpha/\alpha$, even if it might finally turn out to be smaller, will constrain the behavior of the scalar field rather independently of its role for the cosmological acceleration.

It seems nevertheless worth emphasizing that the size of the prediction \reflef{thalp-1}) is not very much away from the observation, allowing us to start with this relation  as a reasonable basis.  The issue is more acute when we notice that the QSO result implies $\dot{\alpha}/\alpha \sim 10^{-16}{\rm y}^{-1}$, and its magnitude, several orders smaller than $t_0^{-1}\sim 10^{-10}{\rm y}^{-1}$, has never been fully understood from a theoretical point of view.

I thank Michael Murphy for providing the QSO data before the publication and reading the manuscript, and also Takashi Ishikawa for the helpful discussion on the error estimates. My thanks are also due to Akira Iwamoto for his crucial comments on the rhenium decay.

\mbox{}\\
\noindent
{\Large\bf References}

\begin{enumerate}
\item\label{ww1}J.K. Webb, V.V. Flambaum, C.W. Churchill, M.J. Drinkwater and J.D. Barrow, Phys. Rev. Lett. {\bf 82}, 884 (1999); M.T. Murphy, J.K. Webb, V.V. Flambaum, V.A. Dzuba, C.W. Churchill, J.X. Prochaska, J.D. Barrow and A.M. Wolfe,
MNRAS, {\bf 327}, 1208 (2001); J.K. Webb, M.T. Murphy, V.V. Flambaum, A. Dzuba, J.D. Barrow, C.W. Churchill, J.X. Prochaska and A.M. Wolfe,  Phys. Rev. Lett. {\bf 87}, 091301 (2001).
\item\label{ww2}M.T. Murphy, J.K. Webb, V.V. Flambaum, 
MNRAS, to be published, astro-ph/0306483.
\item\label{shly}A.I. Shlyakhter, Nature {\bf 264}, 340 (1976).
\item\label{dd}T. Damour and F.J. Dyson, Nucl. Phys. {\bf B480}, 37 (1996).
\item\label{yfetal}Y. Fujii, A. Iwamoto, T. Fukahori, T. Ohnuki, M. Nakagawa, H. Hidaka, Y. Oura and P. M\"{o}ller, Nucl. Phys. {\bf B573}, 377 (2000); Proc. Int. Conf. on Nuclear Data for Science and Technology, 2001, hep-ph/0205206.
\item\label{randp}A.G. Riess {\em et al.}, Astgron. J. {\bf 116}, 1009 (1998); S. Perlmutter {\em et al}., Nature, {\bf 391}, 51 (1998).
\item\label{D98}M.J. Drinkwater {\em et al.}, MNRAS,  {\bf 295}, 457 (1998); C.L. Carilli {\em et al.}, Phys. Rev. Lett. {\bf 85}, 5511 (2000); M.T. Murphy {\em et al.}, MNRAS, {\bf 327}, 1244 (2001).

\item\label{BM1}J.D. Barrow and J. Magueijo, Astrophys. J., {\bf 532}, L87 (2000).
\item\label{BT}J.D. Barrow and C. O'Toole, astro-ph/9904116.

\item\label{wett}C. Wetterich, Phys. Lett. {\bf B561}, 10 (2003).
\item\label{anchor}L. Anchordoqui and H. Goldberg, hep-ph/0306084.
\item\label{gardner}C.L. Gardner, astro-ph/0305080.
\item\label{yf5}Y. Fujii, Phys. Rev. {\bf D62}, 064004 (2000).
\item\label{cup}Y. Fujii and K. Maeda, {\sl The scalar-tensor theory of gravitation}, Cambridge University Press, 2003.

\item\label{bek}J.D. Bekenstein, Phys. Rev. {\bf D25}, 1527 (1982); Phys. Rev. {\bf D66}, 123514 (2002).
\item\label{fon}Y. Fujii, M. Omote and T. Nishioka, Prog. Theor. Phys. {\bf 92}, 521 (1994).
\item\label{bjnporto}Y. Fujii, Proc. First Intn. ASTROD School and Symposium, Beijing, 2001, Int. J. Mod. Phys. {\bf D11}, 1137 (2002), astro-ph/0204069; Proc.  JENAM 2002, Porto, 2002, Astrophysics and Space Science, {\bf 283}, 559 (2003), gr-qc/0212017.
\item\label{olive}K.A. Olive, M. Pospelov, Y.-Z. Qian, A. Coc, M. Cass\'{e} and E. Vangioni-Flam, Phys. Rev. {\bf D66}, 045022 (2002).

\item\label{declmd}Y. Fujii, Phys. Rev. {\bf D26}, 2580 (1982);  D. Dolgov, Proc. Nuffield Workshop, Cambridge University Press, 1982;  L.H. Ford, Phys. Rev. {\bf D35}, 2339 (1987);  Y. Fujii and T. Nishioka, Phys. Rev. {\bf D42}, 361 (1990).
\item\label{yfnn}Y. Fujii, Nature Phys. Sci. {\bf 234}, 5 (1971); Phys. Rev. {\bf D9}, 874 (1974).
\item\label{fsb}E. Fischbach and C. Talmadge, {\sl The search for non-Newtonian gravity}, AIP Press, Springer-Verlag, 1998.
\item\label{eotwash}C.D. Hoyle {\em et al.}, Phys. Rev. Lett. {\bf 86}, 1418 (2001).

\item\label{isochron}M.I. Smoliar, R.J. Walker and J.W. Morgan, Science {\bf 271}, 1099 (1996).
\item\label{yfai}Y. Fujii and A. Iwamoto, in preparation.

\end{enumerate}

\end{document}